\documentstyle[aps,prl,preprint]{revtex}
\newcommand{\be}{\begin{eqnarray}}
\newcommand{\ee}{\end{eqnarray}}

\begin{document}
\draft
\title{Optical Investigations of Charge Gap in Orbital Ordered La$_{1/2}$Sr$_{3/2}$%
MnO$_4$}
\author{J. H. Jung,$^1$ J. S. Ahn,$^1$ T. W. Noh,$^1$ Jinhyoung Lee,$^2$ Jaejun Yu,$%
^1$ Y. Moritomo,$^3$ I. Solovyev,$^4$ and K. Terakura$^4$}
\address{$^1$Department of Physics, Seoul National University, Seoul 151-742, Korea}
\address{$^2$Department of Physics, Sogang University, Seoul 121-742, Korea}
\address{$^3$CIRSE, Nagoya University, Nagoya 464-8603 and PRESTO, JST, Japan}
\address{$^4$ JRCAT, National Institute for Advanced Interdisciplinary Research, Tsukuba, Ibaraki 305, Japan}
\date{\today}
\maketitle

\begin{abstract}
Temperature and polarization dependent electronic structure of La$_{1/2}$Sr$%
_{3/2}$MnO$_4$ were investigated by optical conductivity analyses. With
decreasing temperature, for $E\parallel ab$, a broad mid-infrared (MIR) peak
of La$_{1/2}$Sr$_{3/2}$MnO$_4$ becomes narrower and moves to the higher
frequency, while that of Nd$_{1/2}$Sr$_{3/2}$MnO$_4$ nearly temperature
independent. We showed that the MIR peak in La$_{1/2}$Sr$_{3/2}$MnO$_4$
originates from orbital ordering associated with CE-type magnetic ordering
and that the Jahn-Teller distortion has a significant influence on the width
and the position of the MIR peak.
\end{abstract}


%\twocolumn[\hsize\textwidth\columnwidth\hsize\csname @twocolumnfalse\endcsname

%\vskip1pc]
\newpage

Due to extensive studies of 3{\it d} transition metal oxides, it has been
recognized that correlations among spin, charge, and orbital degrees of
freedom play important roles in their physical properties\cite{imada}.
Especially in doped manganites, such a coupling exhibits very interesting
phenomena: colossal magnetoresistance\cite{s_jin}, magnetic field induced
structural phase transition\cite{asamitsu}, and charge/orbital ordering\cite
{chen}. Recently, much interest has been focused on the charge/orbital
ordering which can be characterized by a real space ordering of Mn$^{3+}$/Mn$%
^{4+}$ ions at a commensurate value of charge carrier, such as 1/8, 1/2, and
2/3. The charge/orbital ordering usually incorporates with a sharp increase
of resistivity, a suppression of magnetic susceptibility, and changes of
lattice constants\cite{kuwahara}.

To get understanding on the charge/orbital ordering, many efforts have been
put into La$_{1/2}$Sr$_{3/2}$MnO$_4$\cite{moritomo95,sternlieb,murakami},
which is known to have a CE-type antiferromagnetic (AFM) ordering below $T_N$
$\sim $ 110 K. Murakami {\it et al.}\cite{murakami} reported diffraction
studies of La$_{1/2}$Sr$_{3/2}$MnO$_4$ using x-ray near the Mn {\it K}%
-absorption edge. From anomalous dispersion of scattering factor for Mn$%
^{3+} $ and Mn$^{4+}$, they claimed that the charge/orbital ordering was
observed directly. To explain why the Mn 3{\it d} orbital ordering can
influence such Mn 1{\it s} $\rightarrow $ 4{\it p} dipole transition,
Ishihara {\it et al.}\cite{ishihara} suggested Coulomb repulsion between the
Mn 3{\it d} and 4{\it p} electrons. However, Elfimov {\it et al}.\cite
{elfimov} pointed out that band structure effects rather than the local
Coulomb repulsion should dominate the polarization dependence of the {\it K}
edge scattering.

In this Letter, we report optical conductivity spectra, $\sigma (\omega )$
of La$_{1/2}$Sr$_{3/2}$MnO$_4$ (LSMO) and Nd$_{1/2}$Sr$_{3/2}$MnO$_4$
(NSMO). Note that the former show an charge/orbital ordering around $Tco$ $%
\sim $ 220 K, but that the latter does not show any ordering at all\cite
{moritomo97}. As temperature ($T$) decreases, a mid-infrared (MIR) peak in
the LSMO ab-plane becomes narrower and a corresponding optical gap
significantly increases. On the contrary, the MIR peak of the NSMO shows
little $T$-dependence. To understand these interesting phenomena, we
calculated the polarization dependent $\sigma (\omega )$ using the
linearized muffin-tin orbital (LMTO) methods and by analysis of the tight
binding (TB) model. The LMTO results were in a remarkable agreement with
experimental ones, indicating that the strong orbital ordering with the
CE-type AFM ordering bring forth the MIR peak. Furthermore, the TB analysis
clearly suggests that the strong $T$-dependences of the optical gap $\Delta $
and the MIR peak of LSMO should be caused cooperatively by the orbital
ordering, the CE-type AFM ordering, and the Jahn-Teller (JT) distortion.

We prepared LSMO and NSMO single crystals using the floating zone methods.
Details of sample growth and characterizations were reported earlier\cite
{moritomo97}. Near normal incident reflectivity spectra $R(\omega )$ were
measured from 0.01 to 6.0 eV with various temperatures and polarizations.
Just before reflectivity measurements, we polished the crystals up to 0.3 $%
\mu $m using diamond pastes. To subtract surface scattering effects, a gold
normalization technique was used. Using the Kramers-Kronig (KK)
transformation, $\sigma (\omega )$ were obtained\cite{jung98}. To reduce
errors of the KK analysis, we also independently measured $\sigma (\omega )$
in the frequency region of 1.5 $\sim $ 5.0 eV using the spectroscopic
ellipsometry (SE). For such optically uniaxial samples, we should measure
ratios of reflectances for {\it p}- and {\it s}- polarized lights at several
incident angles and then calculated optical constants\cite{ellipso}. The SE
results agreed quite well with the KK results, demonstrating the validity of
our KK analysis.

Figures~1(a) and (b) show the polarization dependent $\sigma (\omega )$ of
LSMO and NSMO at 290 K, respectively. Note that the behaviors of $\sigma
(\omega )$ at 290 K are quite similar for both crystals, suggesting the
optical transitions related with the La and the Nd ions should be located at
the energy region higher than 4.0 eV. The $\sigma (\omega )$ in the ab-plane
($E\parallel ab$) are quite different from those along the c-axis ($%
E\parallel c$). [Similar anisotropy could be seen in a bilayer manganite, La$%
_{1.2}$Sr$_{1.8}$Mn$_2$O$_7$\cite{ishikawa98}.] Gap values were estimated
from crossing points of abscissa with linear extrapolations of $\sigma
(\omega )$. For both crystals, $\sigma (\omega )$ for $E\parallel ab$ show
broad peaks around 1.0 and 3.5 eV with $\Delta \sim $ 0.2 eV, and $\sigma
(\omega )$ for $E\parallel c$ show peaks around 1.2 and 4.0 eV with $\Delta
\sim $ 0.7 eV. Since the broad peaks located above 2.0 eV are similar to
those in cubic perovskite manganites\cite{kim98}, these features can be
assigned to O 2{\it p} $\rightarrow $ Mn $e_g$ transitions.

Although $\sigma (\omega )$ for both crystals are very similar at 290 K,
their $T$-dependences are quite different. Figures 2(a) and (b) show $T$%
-dependent $\sigma (\omega )$ of LSMO and NSMO for $E\parallel ab$. [$T$%
-dependences of $\sigma (\omega )$ for $E\parallel c$ are quite small.] For
LSMO, there are large spectral weight changes up to 2.0 eV. With decreasing $%
T$, the spectral weight below 0.8 eV is transferred to a higher energy: $%
\Delta $ increases significantly and the broad peak around 1.0 eV becomes
narrower. For NSMO, there is little $T$-dependence in $\sigma (\omega )$ and 
$\Delta $ is also nearly independent of $T$. From this comparison, we can
argue that the large spectral changes in LSMO should come from the
charge/orbital ordering associated with the CE-type AFM ordering.

To get further insights, we compared our experimental results with
theoretical predictions. Figure 3 shows $\sigma (\omega )$ calculated for Y$%
_{1/2}$Sr$_{3/2}$MnO$_4$ using the LMTO method\cite{igor}. [Even though we
calculated for Y$_{1/2}$Sr$_{3/2}$MnO$_4$, the main features of $\sigma
(\omega )$ are thought to be nearly the same as LSMO.] Since we used the
phenomenological Lorentzian broadening with $\Delta \varepsilon \simeq 0.13$
eV, the value of $\sigma (0)$ is finite even in the insulating state. The
overall features, especially polarization dependence, of the theoretical $%
\sigma (\omega )$ are nearly the same as those in Fig. 1(a). Due to the
limitation of the LMTO method for the higher-energy excitations, the
theoretical value for $\sigma (\omega )$ around 4.0 eV is by a factor of two
smaller than the experimental value. As shown in Fig. 3, the oxygen
displacement $\delta $ between Mn(1) and Mn(2) along the zigzag chain can
induce large spectral weight changes below 2.0 eV.

One of the important issues is what drives the charge/orbital ordering. As
possible candidates, the intersite Coulomb repulsion\cite{satpathy} and the
JT distortion\cite{fujimori} have been considered. Compared to LSMO, NSMO is
known to have a shorter distance of the Mn-O-Mn straight bond\cite
{moritomo97}, which results in a larger intersite Coulomb interaction and a
larger hopping energy of {\it e}$_g$ conduction electrons. However, as shown
in insets of Fig. 2, the measured values of $\Delta $ for LSMO are larger
than those of NSMO. It implies that the conduction electron screening in
NSMO should be dominant, which leads to no magnetic spin ordering. These
results are consistent with recent neutron scattering data which showed no
magnetic ordering in NSMO\cite{mori_neutron} and an AFM\ ordered phase in
the MnO$_2$ layer of LSMO\cite{sternlieb}. Our first principles calculations%
\cite{jinhyoung} revealed that the CE-type AFM ordering produce a strong
orbital ordering even without the JT distortion. Once the orbital ordering
occurs, the JT distortion will be induced. Then it will enhance the orbital
ordering and stabilize the CE-type AFM ordering cooperatively. [Being
consistent with this argument, the magnitude of the JT distortion in the
ab-plane of LSMO is larger than that of NSMO\cite{moritomo97}.]

To clarify effects of the orbital ordering and the JT distortion on the
electronic structure of LSMO, we set up a TB model for the MnO$_2$ plane in
LSMO by taking account of only $e_g$ orbitals at the Mn sites\cite{jinhyoung}%
. [Here, the TB orbital $|e_g\rangle $ should be considered as a Wannier
state, i.e. a superposition of the Mn 3{\it d} and the O 2{\it p} states.]
The model Hamiltonian\cite{millis} can be written as 
\begin{eqnarray}
H &=&\sum_{\langle ij\rangle \alpha \beta \sigma }t_{ij}^{\alpha \beta
}d_{i\alpha \sigma }^{+}d_{j\beta \sigma }-J_H\sum_{i\alpha \sigma \sigma
^{\prime }}\vec{S}_i\cdot \vec{\sigma}_{\sigma \sigma ^{\prime }}d_{i\alpha
\sigma }^{+}d_{i\alpha \sigma ^{\prime }}  \nonumber  \label{eq:tbmodel} \\
&&-g\sum_{i\alpha \beta \sigma }\vec{Q}_i\cdot \vec{\tau}_{\alpha \beta
}d_{i\alpha \sigma }^{+}d_{i\beta \sigma }+\sum_i\frac c2\vec{Q}_i^2,
\end{eqnarray}
where $d_{i\alpha \sigma }$ represents an annihilation operator for the
state at the site $i$ with the orbital index $\alpha $ and spin index $%
\sigma $. It is noted that the $e_g$ states consist of two orbitals, $%
|x^2-y^2\rangle $ and $|3z^2-r^2\rangle $. The second term corresponds to
the Hund coupling of the $e_g$ conduction electrons with the $t_{2g}$
localized spin $\vec{S}_i$ at the site $i$, the third term to the JT type
electron-lattice interaction with the coupling constant $g$, and the last
term to the elastic energy of the JT phonon mode $\vec{Q}=(Q_2,Q_3)$. $\vec{%
\sigma}$ and $\vec{\tau}$ are Pauli matrices. The parameters in the
electronic part of the TB Hamiltonian were determined as $t_{dd\sigma }$ =
0.7 eV, $J_H$ = 0.75 eV, $g$ = 3.85 eV/\AA , and $c$ = 13.58 eV/\AA $^2$\cite
{jinhyoung}.

Assuming the CE-type AFM ordering of the $t_{2g}$ spins, we obtained the
density of state (DOS) without any JT distortion. As shown in Fig.~4(a), DOS
has three separate main peaks, each of which corresponds to bonding (B),
non-bonding (N), and anti-bonding (A) states of the Mn(1) and the Mn(2) $e_g$
orbitals. The B states are fully occupied and separated by the unoccupied N
states with a band gap of $\sim $ 0.2 eV. Due to the peculiar nature of the
1D zigzag chain geometry in the CE-type AFM configuration, the $%
|3x^2-r^2\rangle _1$ orbitals at the Mn(1) sites are strongly hybridized
with the $|x^2-y^2\rangle _2$ components of the $e_g$ orbitals at the
neighboring Mn(2) sites along the chain, while the inter-chain hybridization
is suppressed by the exchange splitting due to AFM coupling. The strong
hybridization along the zigzag chains separate the B and A states by $\sim $
2.0 eV. As a result, the $|3x^2-r^2\rangle _1$ orbital state dominates the
occupancy at the Mn(1) site and leads to the orbital ordered structure in
the MnO$_2$ layer.

The orbital ordered electronic structure together with the JT distortion in
the CE-type AFM state leads to interesting consequences on the interband
transition. While the Mn(1) site maintains its inversion symmetry, the Mn(2)
site at the edge of the zigzag chain has no inversion symmetry due to the
CE-type AFM ordering. Thus, the $e_g$-type Wannier state at the Mn(2) site
becomes a mixture of the $d$- and the $p$-orbital states. Since both types
of Mn atoms are on the mirror plane with respect to the $z$-reflection, no
dipole transition is allowed for $E\parallel c$. On the other hand, in the
case of $E\parallel ab$, the dipole transition at the Mn(2) site becomes
allowed because $\langle {\rm B}_{Mn(2)}|p_{x,y}|{\rm \ N}_{Mn(2)}\rangle
\ne 0$. Therefore, we expect that $\sigma (\omega )$ for $E\parallel c$
should be strongly suppressed below 2.0 eV, while the $\sigma (\omega )$ for 
$E\parallel ab$ have its first peak near 1.0 eV which corresponds to the B $%
\rightarrow $ N interband transition. These TB analyses are consistent with
the experimental result of Fig.~1(a) as well as the LMTO result of Fig.~3.

In Fig.~4(b), we show the joint DOS (JDOS) projected on the Mn(2) site. When
the frequency and polarization dependences of dipole matrix element are
neglected, $\sigma (\omega )$ is considered to be proportional to the JDOS,
since the dipole transition at the Mn(2)\ site without inversion symmetry is
a major contributor. The solid line represents the JDOS without any JT
distortion, and the dashed line with the oxygen distortion of $\delta =0.10$
a.u. We can obtain the JT distortion, $Q_2\approx 3\delta \sqrt{2}$ and $%
Q_3\approx -3\delta \sqrt{6}$ for Mn(1), and $Q_2\approx 0$ and $Q_3\approx
3\delta \sqrt{6}$ for Mn(2) by restricting the volumes of the octahedra
unchanged. The peak near 1.0 eV corresponds to the B $\rightarrow $ N
transition, and the peak near 2.0 eV corresponds to the B $\rightarrow $ A
transition. The overall shape is in close agreement with the LMTO result of
Fig.~3, but the B $\rightarrow $ A feature turns out to be very weak in the
experimental spectrum, shown in Fig.~2(a)\cite{fitting}. In Fig.~4(b), it is
emphasized that the increasing JT distortion results in the narrowing of the
B band and consequently the width of the B $\rightarrow $ N transition as
well, which is quite consistent with experimental observation on the $T$%
-dependence of $\sigma (\omega )$. As $T$ decreases, a fluctuation in the
CE-type AFM ordering is suppressed, the JT distortion increases and the
orbital ordering is enhanced. The observed strong $T$-dependence of $\sigma
(\omega )$ of LSMO is the result of cooperative enhancement of the orbital
ordering.

Even though the $T$-dependence of the MIR peak below $T_{CO}$ in LSMO can be
well understood by the orbital ordered electronic structure with the JT
distortion, the $\sigma (\omega )$ of either LSMO above $T_{CO}$ or NSMO at
all $T$ still exhibit the similar MIR features. Like in the case of Fe$_3$O$%
_4$\cite{anderson}, it could be attributed to the local orbital fluctuation
without a long range CE-type AFM ordering or charge ordering.

In summary, we investigated the orbital ordering in La$_{1/2}$Sr$_{3/2}$MnO$%
_4$ using the optical conductivity analyses. With decreasing temperature,
the peak corresponding to bonding $\rightarrow $ non-bonding transition
shifts to the high frequency and becomes narrower. Comparing with optical
conductivity of Nd$_{1/2}$Sr$_{3/2}$MnO$_4$ and theoretical results, we
conclude that such behaviors could be explained by the CE-type orbital
ordering within the MnO$_2$ layers stabilized by the Jahn-Teller distortion.

This work are financial supported by Ministry of Education through the Basic
Science Research Institute Program No. BSRI-98-2416, by the Korea Science
and Engineering Foundation through RCDAMP of Pusan National University, and
by Ministry of Science and Technology through grant No. I-3-061. This work
was also supported by a Grant-In-Aid for Scientific Research from the
Ministry of Education, Science, Sports and Culture and from Precursory
Research for Embryonic Science and Technology (PRESTO), Japan Science and
Technology Corporation (JST). Partly, this work is also supported by NEDO. 
%%%%%%%%%%%%%%%%%%%%%%%%%%%%%%%%%%%%%%%%%%

\begin{figure}[tbp]
\caption{$\sigma $($\omega $) of (a) LSMO and (b) NSMO for $E\parallel ab$
and $E\parallel c$ at 290 K.}
\label{room}
\end{figure}

\begin{figure}[tbp]
\caption{$T$-dependent $\sigma (\omega )$ for $E\parallel ab$ of (a) LSMO
and (b) NSMO. In the insets of (a) and (b), values of $\Delta $ were also
shown. }
\label{low}
\end{figure}

\begin{figure}[tbp]
\caption{Polarization dependent $\sigma (\omega )$ of Y$_{1/2}$Sr$_{3/2}$MnO$%
_4$ obtained by the LMTO calculation.}
\label{igor}
\end{figure}

\begin{figure}[tbp]
\caption{(a) Energy band diagram for B, N, and A of hybridized Mn(1) and
Mn(2) orbitals. (b) JDOS projected on Mn(2) site without any JT distortion
(solid line) and with JT distortion (dashed line). }
\label{jinhyoung}
\end{figure}


\begin{references}
\bibitem{fitting}  We tried to fit $\sigma $($\omega $) of LSMO up to 4.0 eV
with two Lorentzian or Gaussian functions, but failed. The fitting became
much better with a small peak near 2.0 eV.
\end{references}
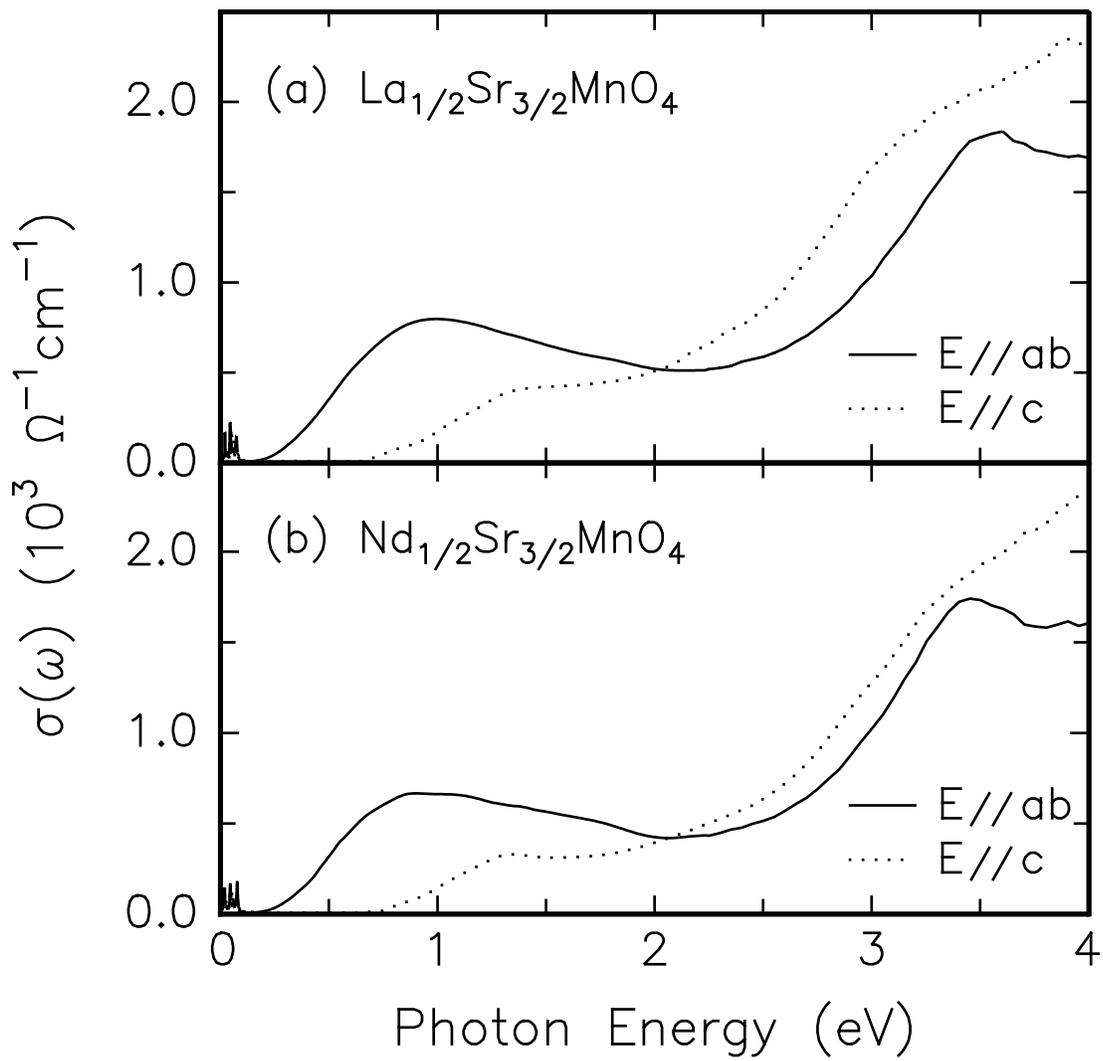
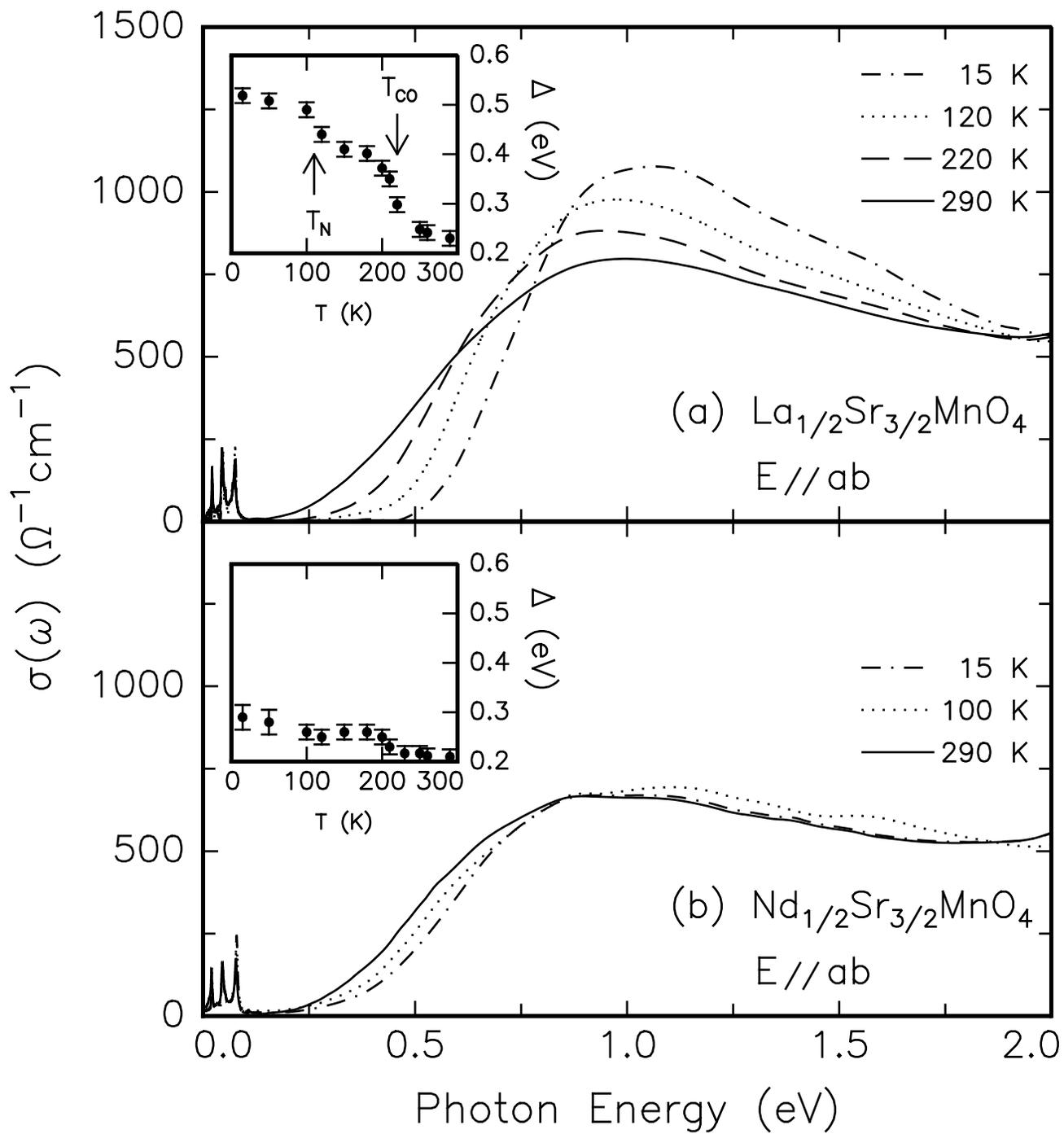
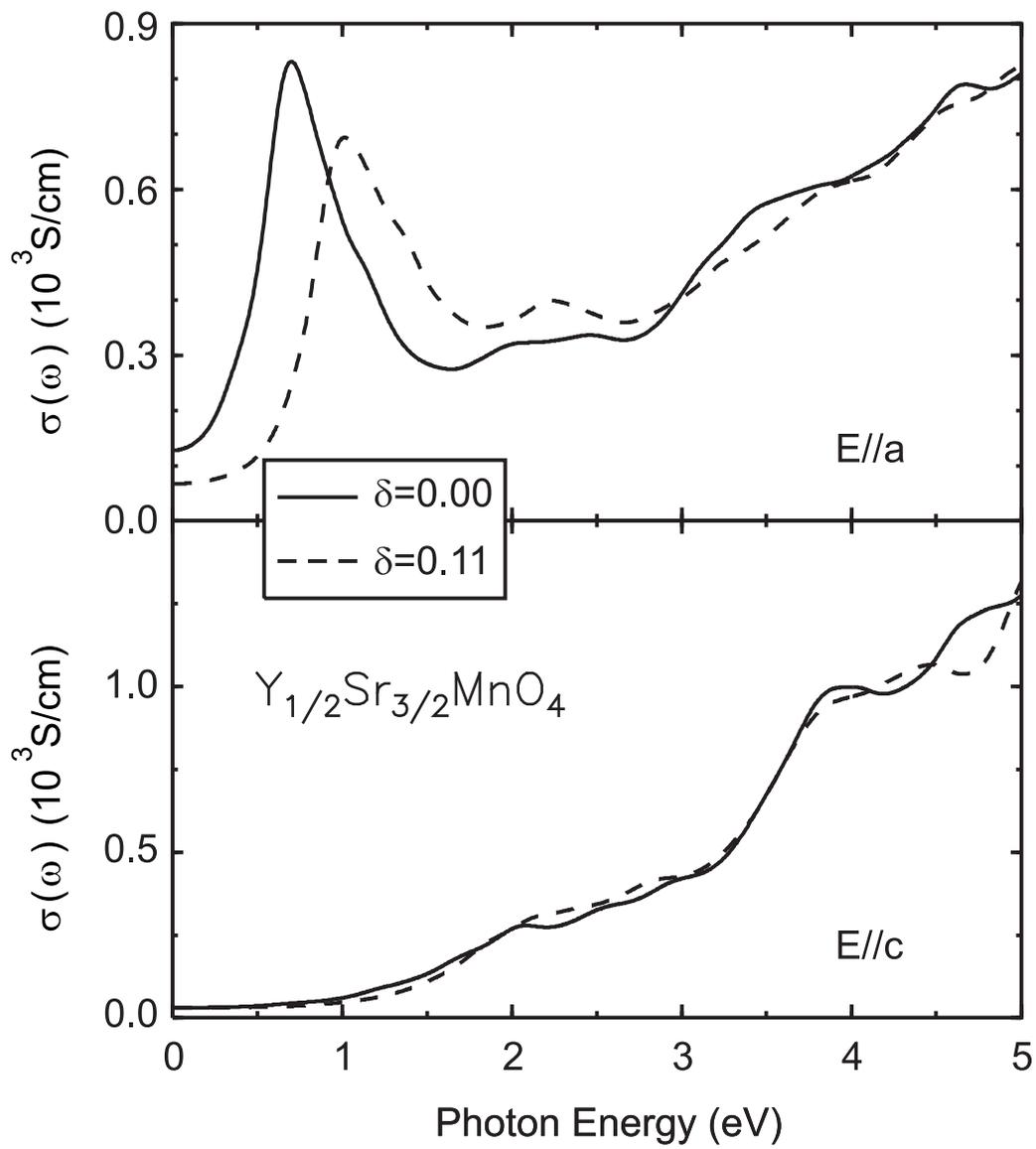
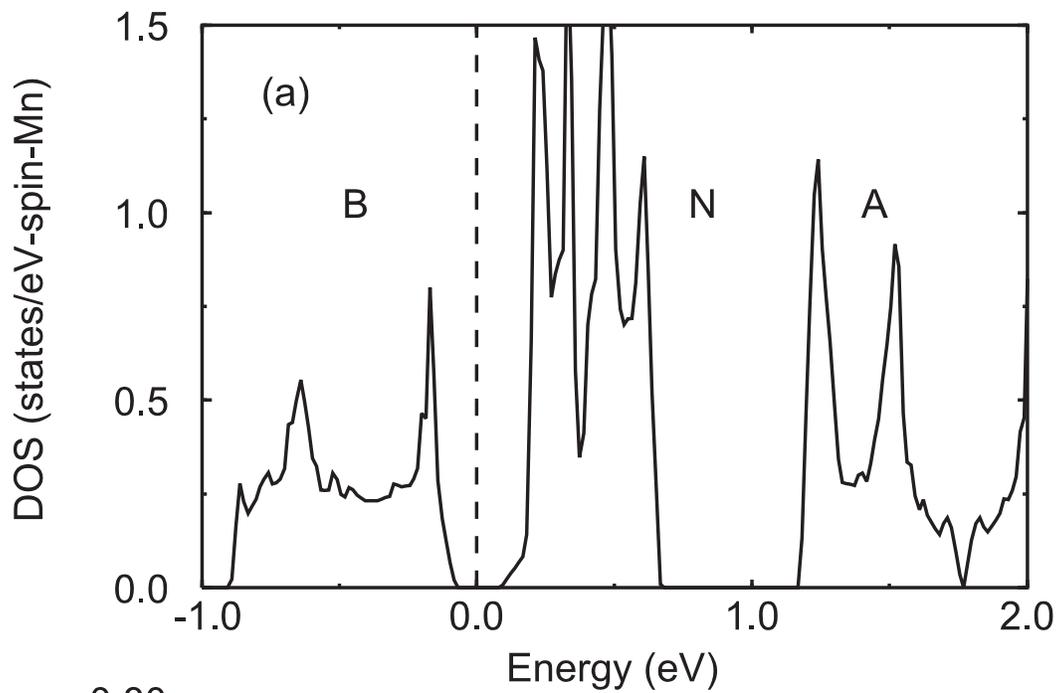
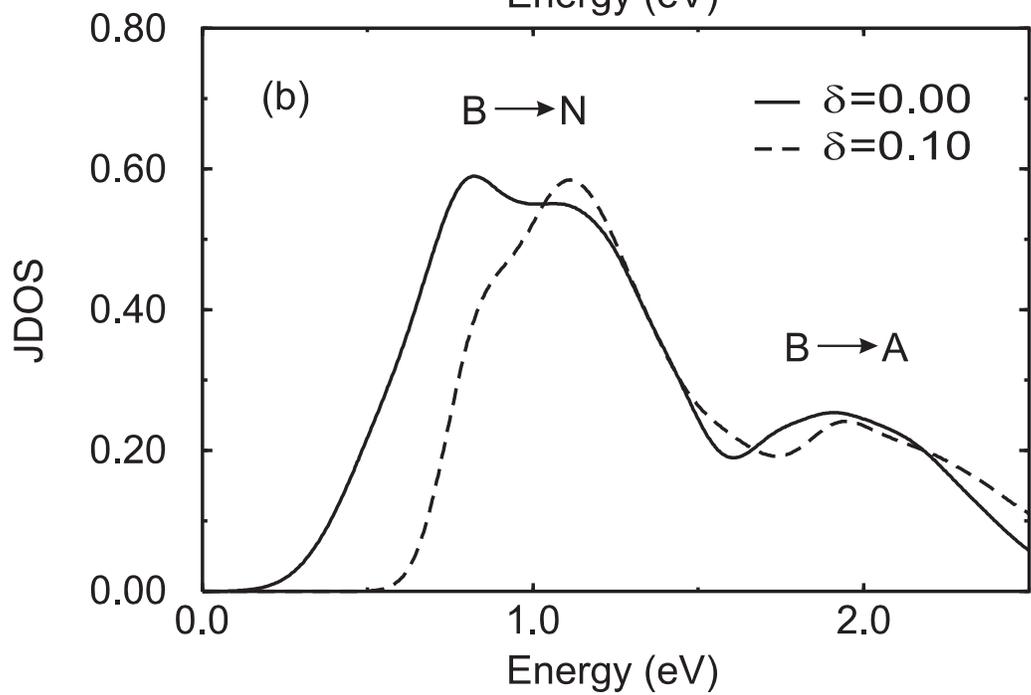
\end{document}